# Venus Topography and Boundary Conditions in 3D General Circulation Modeling

Michael J. Way[1,2] & June Wang[3]

1. Goddard Institute for Space Studies, New York, New York, USA
2. Department of Physics & Astronomy, Uppsala University, Uppsala 75120, Sweden
3. Department of Earth & Planetary Science, Washington University in Saint Louis, St. Louis, MO, 63130, USA

**Abstract.** We describe how one ingests 3D topographic data from NASA's Venus Magellan Spacecraft radar observations into the ROCKE-3D Planetary General Circulation Model. We also explain how boundary condition choices such as ocean/lake coverage/depth, rotation rate, atmospheric constituents and other factors influence surface conditions in ROCKE-3D paleo-Venus simulations. Studies such as these should also be considered when examining liquid water habitability in similar exoplanet experiments.

## 1. Introduction

Given the similar size, density, probable composition (Chassefiere et al. 2012) and proximity of Venus to Earth planetary scientists have struggled to fully understand the differing climatic histories of these two worlds. For example, measurements of the deuterium to hydrogen ratio (D/H) by Pioneer Venus (Donahue et al. 1982, 1997) showed that the hot parched Venus of today had more water in previous epochs. Current estimates are that paleo-Venus could have had the equivalent of between 4 and 525 meters of liquid water if spread across a sphere of radius 6051.85 km (Donahue & Russell 1997). 6051.85 km is the mean radius from Venus as determined from the Magellan mission. How, when and at what rate it lost its water is unclear (e.g. Kasting & Pollack 1983, Donahue 1999). Prior to 2016 most atmospheric modeling of paleo-Venus was via the 1-D models of Grinspoon & Bullock (2007) who found that under the assumption of a 1 bar $N_2$ dominated atmosphere with 6 mb of $H_2O$ at the surface and 100% cloud cover the surface air temperature would be around 27°C. Recent three-dimensional General Circulation Model (3D GCM) work of Way et al. (2016) has shown that within the knowledge constraints of today a paleo-

Venus could have had liquid water on its surface for well over 1 billion years of its early history. Below we describe the input Venus topographic data, the 3D GCM that utilized this data, and the choices made that revealed a paleo-Venus with a habitable climate in its early history.

## 2. Data

The most complete and highest resolution topographic maps of the Venus surface were derived from the NASA Magellan Radar altimetry observations taken from 1990 to 1992 (Ford & Pettengill 1992), together with previous Pioneer Venus (Pettengill et al. 1980) and Venera (Alexandrov et al. 1987, Barsukov et al. 1986) spacecraft observations. The grid resolution of the resulting topographic map is about 1°×1° and the vertical resolution of the data is approximately 80 meters. The topography data are in kilometers with the zero-surface referenced to a sphere of 6051.85 km radius. Different formats of the maps are currently available at the PDS Geosciences node at Washington University in St. Louis[1]. The PDS topographic map of Venus covers a range from -120° to 240° in longitude and from -89.5° to 89.5° in latitude. The map is neither 'centered' at 0°, nor at 180° as is common. The map center was offset by 59.5 degrees to see the Aphrodite Terra feature unobstructed by the edges of the display frame. Aphrodite Terra is located near the equator and includes four smaller highlands: Ovda, Thetis, Atla, and Ulfrun Regiones (Young 1990).

## 3. Method

The topographic data from Section 2 was utilized in a 3D GCM known as ROCKE-3D (Way et al. 2017, hereafter Way17). Historically 3D GCMs (also described as Global Climate Models) were constructed to study Earth climate over relatively long periods of time (100s of years) in comparison with higher resolution models that studied weather phenomena on short time scales (less than 1 month). Today's Earth GCMs are coupled atmosphere-ocean models with detailed physics of radiative transfer, sea ice, land ice, ground hydrology, cloud dynamics and a host of other processes that govern the climate of Earth (e.g. Schmidt et al. 2014). They have been instrumental in understanding the possible impacts of climate change on Earth.[2] However, planetary scientists quickly realized that the same models could be used to model other planetary environments. Leovy & Mintz (1969) were the first to model another planet (Mars) and Joshi et al. (1997) made the first exoplanet GCM experiments. Presently there are a host of GCMs modeling the solar system planets, dwarf planets, and moons with atmospheres in our solar system and hypothetical exoplanetary systems. Examples can be found in the Introduction section of Way17.

---

[1] http://pds-geosciences.wustl.edu/mgn/mgn-v-rss-5-gravity-l2-v1/mg_5201/

[2] http://www.ipcc.ch/

The topography data described in Section 2 above were ingested into a NetCDF [3] compatible format (NetCDF 2016) where the longitudinal offset mentioned above was taken into account so that zero longitude begins on the left-hand side as is standard in ROCKE-3D when displayed. In the specific case of the paleo-Venus simulations related to Way et al. 2016 (hereafter Way16) the data were down-sampled from the original resolution of 1°× 1° to 4°×5° in latitude and longitude. Compared to modern climate simulations with the ROCKE-3D parent model GCM ModelE2 (Schmidt et al. 2014), resolution of 4°×5° may seem quite coarse, but there are two reasons why Way16 chose this resolution. 1.) The model run-time is a factor of 2–3 faster at this lower resolution compared to ModelE2 which runs at the resolution of 2°×2.5° for the atmosphere and 1°×1.25° for the ocean. 2.) The lower resolution has little effect on the climate dynamics of a slowly rotating (day length greater than 64 Earth days long, see Way et al. 2015) world like Venus. This is because the Rossby radius of deformation is much larger than the 4°×5° grid box size (Del Genio & Suozzo 1987).

In Way16 an ocean depth compatible with the D/H measurements mentioned above was desired for the simulations. The mean planetary radius of 6051.85 km was chosen as the ocean top. In the source topographic input files from the PDS any down-sampled grid cell altitude with a negative 'height' would be assigned as an ocean cell. Any cell with a positive value would be assigned as a land grid cell. The resultant ocean had an average depth of 311 meters, if spread across a hypothetical flat surface sphere of radius 6051.85 km. This resulting ocean volume is within the estimated D/H range of possibilities in Donahue & Russell (1997). The result landmask topography is shown in Figure 1. Additionally, another step was taken: to prevent any possibility of issues related to particular ocean grid cells freezing to the bottom[4] and to simplify interpretation of land climatic effects, a number of small land grid cells with surrounding shallow ocean depths were transformed to sea grid cells with a depth set to the average of the deeper surrounding grid cells. One can see the elimination of these land grid cells when looking at Figure 1a (with small islands) versus 1b (without small islands). The higher resolution map from the PDS was further utilized to calculate a roughness length in each 4°×5° land grid cell based on the standard deviation of the twenty 1°×1° cells used to construct each 4°×5° grid cell. A water run-off direction was calculated based on the land topography so that lakes and rivers will form where necessary. For example, at points where there is a local minimum a lake will be allowed to form if precipitation outweighs evaporation on the same timescale. Lakes are given as fractional amounts of a given grid cell, from 0% to 100%. Bare soil of a 50/50 mix of Clay/Sand and an albedo of 0.2 was chosen following Yang et al. 2014.

---

[3] http://www.unidata.ucar.edu/software/netcdf

[4] A known issue with ROCKE-3D for shallow ocean grid cells is that if they freeze to the bottom the GCM will crash as it is incapable of changing topographically defined surface types (e.g. ocean -> land or vice versa).

It is believed that Venus has been largely resurfaced the past several hundred million years (McKinnon et al. 1997, Kreslavsky et al. 2015). Hence today's topographic relief may not be an accurate reflection of its ancient topography, although Way16 made the assumption that the resultant topography was the same over the epoch of the simulation: from 2.9 Gya to 0.715 Gya. The 2.9Gya was chosen to avoid early solar UV activity and geothermal heating effects that are hard to constrain with present day knowledge. The 0.715Gya was chosen as a period before the observed global resurfacing event. We were also able to obtain spectra from Mark Clare (Claire et al. 2012) for these particular times, hence the precision in the dates given. Typically, exoplanetary GCM work has relied upon either Earth topographic relief, idealized continental structures or assumed aquaplanet[5] surfaces (Yang et al. 2014). Way16 took the approach that using the present-day topography of Venus is the least-worst choice for a paleo-Venus world. However, comparable simulations in Way16 were performed with a present-day Earth-like topography. The main differences in these two topographies (as described in Way16) were the ratio of land-to-sea and ocean volume. The final paleo-Venus reconstruction used in Way16 has 10% of the water volume of modern Earth. The Venus topography also has more low-latitude land distribution versus the high-latitude land distribution of modern Earth. These differences played out in the final globally averaged surface temperatures ($T_{surf}$) where the Earth-like topography had a value of $T_{surf}$=23°C. This was 12°C warmer than the Venus topography value of $T_{surf}$=11°C (see Table 1 in Way16) for the same incident solar flux. This is likely because of differing latitudinal distributions of land/sea & oceanic currents. Since Way16 we have explored other possible topographies. Aquaplanet simulations have been carried out assuming an ocean covered surface with a depth of 899 meters (to be published in a forthcoming paper). Those simulations have $T_{surf}$=31°C. This is higher than that of either the Venus or Earth topographic land reliefs mentioned above. This is due to the fact that no land ice is allowed to form at the poles, nor on the dark side of the slowly rotating planet at higher altitudes, as seen in the other topo-graphic reliefs in Way16. These cold land surfaces tend to lower $T_{surf}$. Because of efficient ocean heat transport, there is no ocean ice anywhere in the aquaplanet simulation, hence the higher $T_{surf}$=31°C. One other topography attempted was a land planet simulation where 10 meter equivalent depth water was spread over the Venus topography in lakes or small seas. This unpublished simulation has $T_{surf}$=21°C.

---

[5] A surface covered in 100% water.

Several other key assumptions were made in the modeling efforts regardless of topography: 1.) paleo-Venus would have the same orbital characteristics (rotation, obliquity, eccentricity) as present-day Venus, 2.) the solar insolation and spectral distribution from several solar epochs were obtained from the models of Claire et al. (2012) and were ingested into ROCKE-3D: 2.9Gya, 0.715Gya, present day, and 3.) the atmospheric pressure and constituents would be similar to modern Earth with a 1013mb $N_2$ dominated atmosphere with $CO_2$=400ppmv, $CH_4$=1ppmv and an $H_2O$ profile similar to the 6mb one of Grinspoon & Bullock (2007) (see plate 1 in their paper), but no aerosols, $O_2$ nor $O_3$.

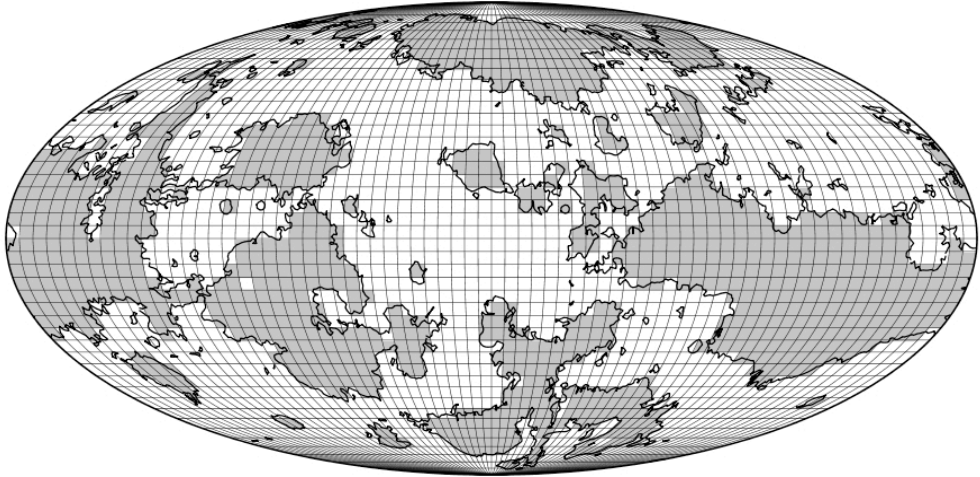

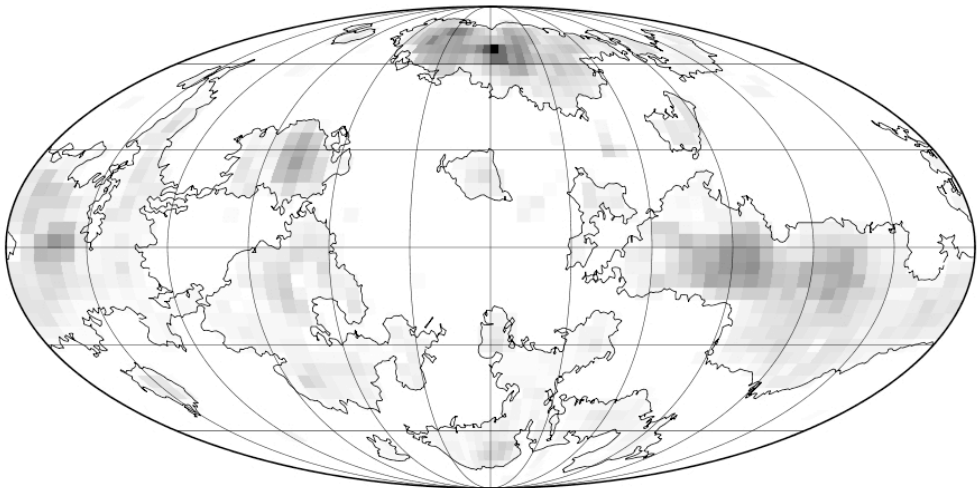

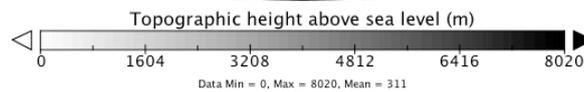

**Figure 1.** Top (1a): Reconstructed topography from Magellan spacecraft data. Grid cells shown are equivalent to the 4°×5° latitude longitude resolution of the NetCDF ROCKE-3D GCM input file. Dark Gray=land, light gray=ocean. Bottom (1b): Modified reconstructed topography from the Magellan spacecraft radar observations. Small islands have been removed. White=ocean, gray gradient=topographic height from 1 (light gray) to 8020 meters (black). The center of the image's 30° spaced grid lines corresponds with 0° latitude and longitude. Each grid line to the north/south is +/- 30° of latitude and to the east/west is +/- 30° of longitude. West is to the left and east is to the right.

Input and output NetCDF (2016) formatted diagnostic data from the GCM are typically displayed using a widely available program called Panoply.[6] When viewing GCM diagnostic output it is convenient to have an 'overlay' that outlines the boundaries of the continents, inland seas, and islands so that one knows if, for example, the surface humidity or temperature you are viewing is over land or sea. This is in contrast to contour lines that are often used to indicate the height of topographic features. The process to generate overlays involves loading the land/ocean topography map into NCAR Graphics[7] where one can then draw the desired boundaries with a mouse. A latitude x longitude sequence of points that comprises each polygon is then saved in a text file. For Panoply, each collection of latitude x longitude points that denote an island or continent are required by Panoply to be delimited by 999999 (an arbitrarily chosen number by the authors of Panoply). This file is then loaded into Panoply and shows up as an optional overlay. The Panoply Venus overlay is shown in the Mollweide projection of Figure 1.

In addition to the definition of land and ocean mentioned above the input topography files also require the definition of lakes if present, or land snow/ice (e.g. Antarctica). Way16's paleo-Venus simulations assumed there was no land snow/ice nor were lakes included at model start. A general approach to defining where lakes might be located involves the use of a 'river directions' file. This ROCKE-3D input file defines the directions where rivers would flow between grid cells based on their relative elevations in the topography file. The downstream flow is always to the grid cell with the greatest relative lower elevation. Most of the rivers will exit into the oceans. If a grid cell resides in a valley where all neighboring grid cells are at higher elevations, a lake may form. Lakes in ROCKE-3D can be variable quantities such that defined lakes in the topography input file can later evaporate, or form elsewhere as dictated by precipitation, evaporation and topography. Figure 3a shows an example of this for the Venus land planet mentioned above that started with 10m depth water equivalent if spread across the surface. Figure 3b shows what this hypothetical land Venus world would look like after running the model 4000 years. Looking at ground hydrology diagnostic outputs we can estimate when they have come into equilibrium.

---

[6] http://www.giss.nasa.gov/tools/panoply

[7] http://ngwww.ucar.edu/

In this case it was 4000 years of model run time. There are also 'emergency' river directions. These may be invoked when a given grid cell has no defined river direction because it was surrounded by higher elevation grid cells, but the lake height in this grid cell increases (due to precipitation and a lack of evaporation) to an elevation higher than one of the surrounding grid cells. When this occurs the emergency river direction is invoked and water is allowed to flow into the (now) lower neighboring grid cell.

Completed GCM runs are normally analyzed or post-processed by examining quantities such as surface temperature, wind speed, albedo, etc. Using Panoply a large selection of projections are also available in addition to the Mollweide used in Figure 1. In Figure 2 we present results from one of the paleo-Venus aquaplanet runs using a Robinson projection. The figures are the results of averaging over 1/6 of a diurnal cycle (approximately 19 earth days in length). Figure 2a is a typical GCM diagnostic output called high level cloudiness (known as PCLDH within ROCKE-3D). Because of the very slow rotation rate of this world the atmospheric dynamics dictate that it has a single large Hadley type cell in the northern hemisphere and a corresponding one in the south. On Earth (a fast spinner) the Coriolis force acts to break up the circulation pattern into several northern and southern Hadley cells. In Figure 2a the sub-stellar point is located at the center of the maximum of PCLDH. These high-level clouds also have a very high albedo and hence this map is not only a PCLDH map, but would also mirror the albedo map if we had shown it. It is these clouds, their high albedo and their contiguous nature that prevent a paleo-Venus type world at 0.715 Gya from having extremely high surface temperatures even though it sees almost 1.7 times as much solar flux as modern-day Earth (see Table 1 in Way16). Figure 2b shows the ground level wind speeds. One can see that the ground level winds below the peak of the high-level clouds near the equator have very high relative velocities. These winds drive the large convective updraft that creates the high-level clouds and borders each hemisphere's Hadley cell.

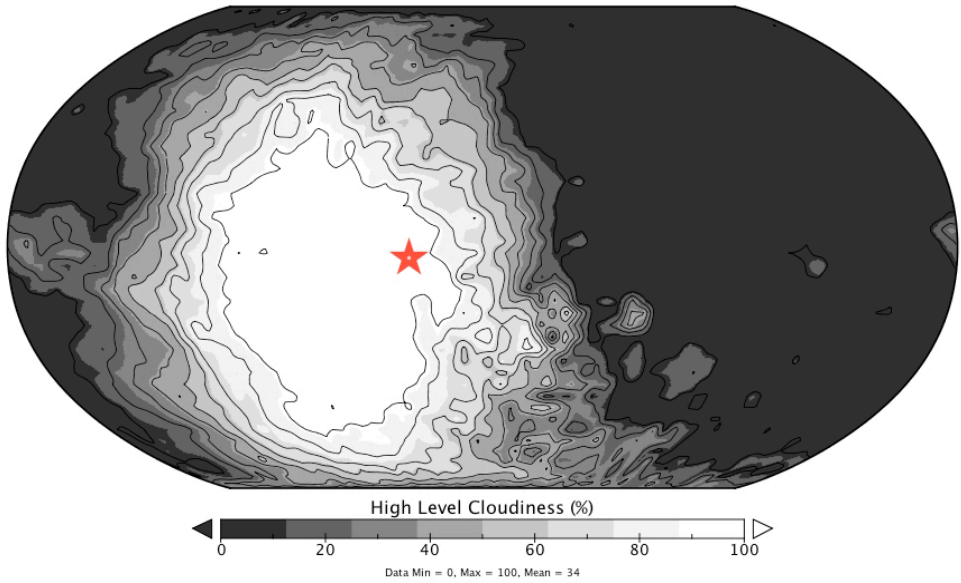

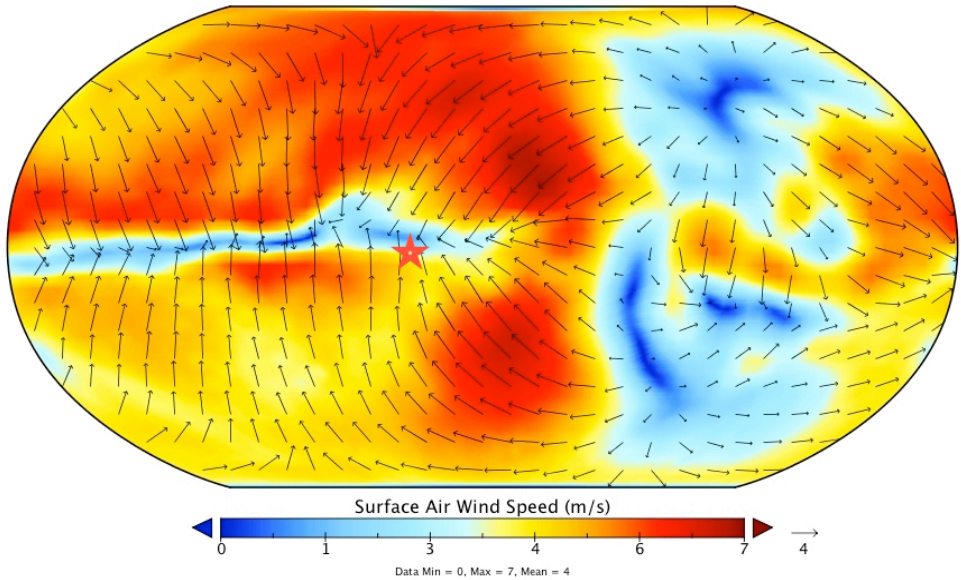

Figure 2: ROCKE-3D aquaplanet simulations of paleo-Venus at 0.715Gya with a constant 899m depth fully-coupled ocean. Red star in each is the location of the substellar point. Each image is an average over 1/6 of a diurnal cycle (19 Earth days in length) and they are taken from the same exact

time in the simulation. Top (2a): Black=0% PCLDH, White=100% PCLDH. Note that the substellar point is not at the center of the high albedo clouds because the planet is rotating and the clouds are an average over 1/6th of a diurnal cycle. Bottom (2b): Surface wind speeds where the longest vectors represent speeds up to 7m/s. These maps were generated in Panoply from post-processed ROCKE-3D data.

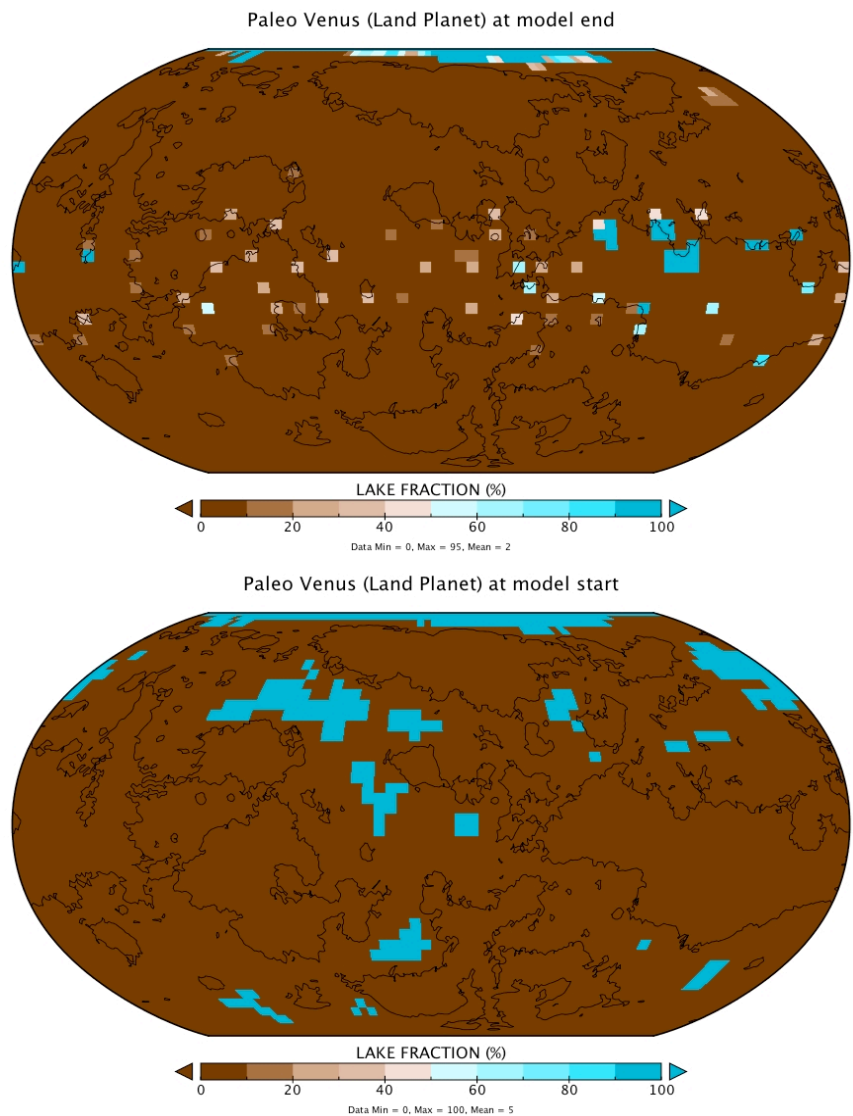

Figure 3: ROCKE-3D land planet simulations of paleo-Venus at 2.9 Gya with 10 m equivalent depth of water if spread across the surface of a sphere

of radius 6051.85 km. Top (3a) shows the lakes residing at the lowest points in the topography at model start. Bottom (3b) shows where the lakes are after running 4000 model years. One can see that precipitation and evaporation patters in the atmosphere have tended to consolidate the lakes into larger contiguous regions. This simulation also used a 50/50 sand/clay soil type.

## 4. Conclusions

The process to transform Magellan spacecraft topographic data into useful input files at the default resolution of 4°×5° for the ROCKE-3D GCM is described. The horizontal and vertical resolution of the Magellan topographic data are sufficient for exploring Venus' climatic history with modern day GCMs (Way et al. 2016), but higher resolution data are desired as modern GCMs can utilize such data. Unfortunately, higher resolution data can only be obtained via another mission to Venus, and currently no space agency has such a mission presently funded. The ROCKE-3D GCM topographic NETCDF input and output files may be obtained from the authors upon request.

### Acknowledgements

This research was supported by the NASA Astrobiology Program through our participation in the Nexus for Exoplanet System Science, and by the NASA Planetary Atmospheres Program, Exobiology Program, and Habitable Worlds Program. We also acknowledge internal Goddard Space Flight Center Science Task Group funding. Special thanks to Jeff Jonas, Robert Schmunk, and Linda Sohl for advice related to the creation of the topographic overlay file.